\documentstyle[12pt]{article}

\def\baselinestretch{1.2}
\def\href#1#2{#2}  

\newcommand{\norm}[1]{\raise.3ex\hbox{:} #1 \raise.3ex\hbox{:}\,}

\def\T{{\cal T}}
\def\O{{\cal O}}
\def\tr{{\rm Tr}}
\def\det{{\rm det}}
\def\MeV{\ {{\rm MeV}}}
\def\GeV{\ {{\rm GeV}}}
\newcommand{\beq}{\begin{equation}}
\newcommand{\eeq}{\end{equation}}
\def\stop{.}
\def\comma{,}
\newcommand{\beqar}{\begin{eqnarray}}
\newcommand{\eeqar}{\end{eqnarray}}
\def\non{\nonumber}
\def\appendix{{\newpage\section*{Appendix}}\let\appendix\section%
        {\setcounter{section}{0}
        \gdef\thesection{\Alph{section}}}\section}

\catcode`\@=11

\@addtoreset{equation}{section}
\catcode`\@=12

\def\noj#1,#2,{{\bf #1} (19#2)\ }
\def\jou#1,#2,#3,{{\sl #1\/ }{\bf #2} (19#3)\ }
\def\ann#1,#2,{{\sl Ann.\ Physics\/ }{\bf #1} (19#2)\ }
\def\cmp#1,#2,{{\sl Comm.\ Math.\ Phys.\/ }{\bf #1} (19#2)\ }
\def\ma#1,#2,{{\sl Math.\ Ann.\/ }{\bf #1} (19#2)\ }
\def\jd#1,#2,{{\sl J.\ Diff.\ Geom.\/ }{\bf #1} (19#2)\ }
\def\invm#1,#2,{{\sl Invent.\ Math.\/ }{\bf #1} (19#2)\ }
\def\cq#1,#2,{{\sl Class.\ Quantum Grav.\/ }{\bf #1} (19#2)\ }
\def\cqg#1,#2,{{\sl Class.\ Quantum Grav.\/ }{\bf #1} (19#2)\ }
\def\ijmp#1,#2,{{\sl Int.\ J.\ Mod.\ Phys.\/ }{\bf A#1} (19#2)\ }
\def\jmphy#1,#2,{{\sl J.\ Geom.\ Phys.\/ }{\bf #1} (19#2)\ }
\def\jams#1,#2,{{\sl J.\ Amer.\ Math.\ Soc.\/ }{\bf #1} (19#2)\ }
\def\grg#1,#2,{{\sl Gen.\ Rel.\ Grav.\/ }{\bf #1} (19#2)\ }
\def\mpl#1,#2,{{\sl Mod.\ Phys.\ Lett.\/ }{\bf A#1} (19#2)\ }
\def\nc#1,#2,{{\sl Nuovo Cim.\/ }{\bf #1} (19#2)\ }
\def\np#1,#2,{{\sl Nucl.\ Phys.\/ }{\bf B#1} (19#2)\ }
\def\pl#1,#2,{{\sl Phys.\ Lett.\/ }{\bf #1B} (19#2)\ }
\def\pla#1,#2,{{\sl Phys.\ Lett.\/ }{\bf #1A} (19#2)\ }
\def\pr#1,#2,{{\sl Phys.\ Rev.\/ }{\bf #1} (19#2)\ }
\def\prd#1,#2,{{\sl Phys.\ Rev.\/ }{\bf D#1} (19#2)\ }
\def\prl#1,#2,{{\sl Phys.\ Rev.\ Lett.\/ }{\bf #1} (19#2)\ }
\def\prp#1,#2,{{\sl Phys.\ Rept.\/ }{\bf #1C} (19#2)\ }
\def\ptp#1,#2,{{\sl Prog.\ Theor.\ Phys.\/ }{\bf #1} (19#2)\ }
\def\ptpsup#1,#2,{{\sl Prog.\ Theor.\ Phys.\/ Suppl.\/ }{\bf #1}
(19#2)\ }
\def\rmp#1,#2,{{\sl Rev.\ Mod.\ Phys.\/ }{\bf #1} (19#2)\ }
\def\yadfiz#1,#2,#3[#4,#5]{{\sl Yad.\ Fiz.\/ }{\bf #1} (19#2) #3%
\ [{\sl Sov.\ J.\ Nucl.\ Phys.\/ }{\bf #4} (19#2) #5]}
\def\zh#1,#2,#3[#4,#5]{{\sl Zh..\ Exp.\ Theor.\ Fiz.\/ }{\bf #1}
(19#2) #3%
\ [{\sl Sov.\ Phys.\ JETP\/ }{\bf #4} (19#2) #5]}

\begin{document}

\begin{titlepage}

\begin{flushright}
NSF-ITP-98-089\\
CERN-TH/98-298\\
hep-th/9809106
\end{flushright}
\vfil\vfil

\begin{center}

{\Large {\bf Aspects of QCD Dynamics from  String Theory}}

\vfil

Akikazu Hashimoto$^a$  and Yaron Oz$^b$

\vfil

$^a$Institute for Theoretical Physics\\
 University of California\\
 Santa Barbara, CA  93106

\vfil

$^b$Theory Division, CERN\\
CH-1211, Geneva 23, Switzerland
\end{center}

\vspace{5mm}

\begin{abstract}
\noindent 
We study dynamical aspects of large $N$ QCD$_4$ via supergravity on
Anti de Sitter black hole geometry.  We compute the mass spectrum, the
topological susceptibility and the gluon condensate and make a
comparison to lattice simulations.  We present analogous results for
QCD$_3$ .
\end{abstract}

\vfil\vfil\vfil
\begin{flushleft}
September 1998
\end{flushleft}
\end{titlepage}

\newpage
\renewcommand{\baselinestretch}{1.05}  

\section{Introduction}

Maldacena \cite{mal} conjectured the existence of a duality between
large $N$ superconformal field theories and superstring or M theory on
Anti de Sitter (AdS) backgrounds with R-R fluxes.  This conjecture
reveals interesting properties of the superconformal field theories,
in particular in the strong coupling region.  In
\cite{WittenAdsThermal} Witten proposed an extension of the duality to
non-supersymmetric theories such as pure QCD.  The idea was to heat
the superconformal field theories and break the superconformal
invariance.  The AdS space is then replaced by the Schwarzschild
geometry describing a black hole in the AdS space.

When the curvature of the space is small compared to the string scale
and the Planck scale, supergravity provides an adequate effective
description of string and M theory respectively. This description has
been extensively studied recently.  It exhibits qualitative agreements
with pure QCD in three and four dimensions \cite{GO98}. In particular
confinement and mass gap are two of the important properties of the
supergravity description \cite{WittenAdsThermal}.  This leads to the
hope that we are at least describing a theory which is in the same
universality class as QCD.

One can also perform quantitative computations via the supergravity
description.  For instance, the spectrum of the scalar glueball masses
has been computed resulting in a surprising agreement with lattice
simulations \cite{COOT98}.  The supergravity limit of string theory is
the limit where the string tension is infinite.  This corresponds to
the strong coupling limit of the gauge theory.  The strong coupling
expansion in the gauge theory is the $\alpha'$ expansion of string
theory. It is therefore clear that a full solution of QCD requires a
detailed knowledge of this expansion, which means an understanding of
string theory with R-R fields on these backgrounds.  Since
supergravity contributes the leading term in this expansion it is
important to have a detailed understanding of it.  We will take this
viewpoint in this paper and use the supergravity description to study
various aspects of QCD.  Note, however, that an important assumption
that is made is that there is no phase transition as a function of the
't Hooft parameter $g_{YM}^2 N$.

The paper is organized as follows:

\noindent In section 2, we describe in detail the supergravity
background and the conjectured correspondence to QCD$_4$ .

\noindent In section 3 we compute glueball mass spectrum.

\noindent In section 4 we compute the topological susceptibility.

\noindent In section 5 we compute the gluon condensate.

\noindent In section 6 we make a detailed comparison with lattice and
field theory results.

\noindent Section 7 is devoted to a discussion and conclusions where
we also present analogous results for QCD$_3$ .

\section{QCD Via String Theory}

The starting point for the construction of QCD$_4$ is the
superconformal theory in six dimensions realized on $N$ parallel
coinciding M$5$-branes.  Compactification of the theory on a circle of
radius $R_1$ gives a five-dimensional theory whose low-energy
effective theory is the maximally supersymmetric $SU(N)$ gauge theory
with gauge coupling constant $g_5^2 = 2 \pi R_1$.  In order to
describe QCD$_4$, one compactifies this theory further on another
${\bf S}^1$ of radius $R_2$. The gauge coupling constant $g_{YM}$ in
four dimensions is given by $g_{YM}^2 = g_5^2/R_2 = 2 \pi R_1/R_2$. In
order to break supersymmetry, one imposes the anti-periodic boundary
condition on the fermions around the second ${\bf S}^1$.

As conjectured in \cite{mal}, the large $N$ limit of the
six-dimensional theory is $M$ theory on AdS$_7 \times {\bf S}^4$. Upon
compactification on ${\bf S}^1 \times {\bf S}^1$ and imposing the
anti-periodic boundary conditions around the second ${\bf S}^1$, we
find $M$ theory to be on the black hole geometry
\cite{WittenAdsThermal}.  Taking the large $N$ limit while keeping the
't Hooft parameter $2 \pi \lambda = g_{YM}^2 N$ finite requires taking
$R_1 \ll R_2$. In this limit, $M$ theory reduces to type IIA string
theory and the M$5$ brane wrapping on ${\bf S}^1$ of radius $R_1$
becomes a D$4$ brane, and the large $N$ limit of QCD$_4$ becomes
string theory on a black hole geometry.

\subsection{The Supergravity Background}

The action for the eleven dimensional supergravity is \footnote{The
$N=1$ supergravity action in eleven dimensions also contains
Chern-Simons terms and a boundary extrinsic curvature terms which we
omitted.}
\beq
S = {1 \over 2 \kappa_{11}^2} \int d^{11}x
\sqrt{g} \left[ R  + {1 \over 48} H^2 \right]
\stop
\label{11d}
\eeq
The reduction of (\ref{11d}) to ten dimensions on a circle of period
$\frac{\lambda}{NT}$ gives
\beq
S = {1 \over 2 \kappa_{10}^2} \int d^{10}x
\sqrt{g} \left[ e^{-2 \phi}\left(R + 4 (\nabla \phi)^2 \right) + {1 \over 4} G^2 + {1 \over 48} H^2 \right] ,
\label{10d}
\eeq
where
\beq
\kappa_{10}^2 = (N/\lambda) T\kappa_{11}^2
\stop
\label{conv1}
\eeq
           
The near horizon limit of the black M5-brane metric is
given by \cite{DPL96,GKT98}
\begin{eqnarray}
ds^2 &=& {y^2 \over L^2}\left[(1-{y_0^6 \over y^6}) dt^2 + \sum_{i=1}^5
dx_i^2 \right] + {4 L^2 \over y^2}(1 - {y_0^6 \over y^6})^{-1} dy^2 +
L^2 d \Omega_4^2
\comma \\
\label{metric2}
H & = & 3 \sqrt{6} L^3  d\Omega_4
\end{eqnarray}
where
\beq
L^9 = N^3 {\kappa_{11}^2 \over 2^7 \pi^5}, \qquad y_0 = {4 \pi L^2 T
\over 3}
\stop
\label{conv}
\eeq
and $d\Omega_4$ is the volume form on the 4-sphere.  Compactifying
$x^5$ on a circle and using the M theory/Type IIA duality one gets the
Type IIA background corresponding to a non extremal D4 brane with
worldvolume coordinates $x^1,..,x^4,t$
\beq ds^2 = {y^3 \over L^3}\left[(1-{y_0^6 \over y^6}) dt^2 +
\sum_{i=1}^4 dx_i^2 \right] + {4 L \over y}(1 - {y_0^6 \over
y^6})^{-1}\, dy^2 + L y\, d \Omega_4^2 \comma \qquad e^{\phi} =
\left({ y \over L } \right)^{3\over 2} \comma \eeq
where $\phi$ is the dilaton field.

\subsection{Fluctuations}

As will be discussed in the next subsection, in order to compute the
correlation functions of QCD operators using supergravity/gauge theory
correspondence, we need to examine the dynamics of small fluctuations
of the supergravity fields. We will be interested in scalar operators
of QCD$_4$.  The relevant fields of Type IIA supergravity are the
dilaton $\phi$, the volume factor, and the time component of the RR
1-form $C_0$. There will be a mixing between the type IIA dilaton and the
volume factor which must be disentangled. A useful parameterization of
fluctuations $f(x_1,y)$, $g(x_1,y)$, and $h(x_1,y)$, which in the language of
eleven dimensional supergravity is appropriately diagonal, is
\begin{eqnarray}
ds^2 &=& 
\left (1-{4 \over 5} g(x_1,y)\right) \sum_{i=0}^1 \dot{g}_{ii} dx^i dx^i
+ 
\left(1- {2 \over 5} f(x_1,y) -{2 \over 5} g(x_1,y) \right) \sum_{i=2}^4 \dot{g}_{ii} dx^i dx^i \nonumber \\
&& + 
\left(1+ {6 \over 5} f(x_1,y) -{4 \over 5} g(x_1,y)\right)
 \dot{g}_{55} \left( dx^5 + C_i dx^i\right)^2 \nonumber \\
&& + \left(1-{4 \over 5} g(x_1,y) \right) \dot{g}_{yy} dy^2
+ (1+ g(x_1,y)) \dot{g}_{\Omega \Omega } d\Omega^2
\end{eqnarray}
where $\dot{g}$ is the background metric.  We consider the case
$C_0=h(x_1,y)$ and $C_{i\ne 0} =0$. Without loss of generality,
$f(x_1,y)$, $g(x_1,y)$, and $h(x_1,y)$ are assumed to depend only on
$x_1$ and $y$. In the language of Type IIA supergravity, these
fluctuations are parameterized according to
\begin{eqnarray}
ds^2 &=& 
\left (1\!+\!{3 \over 5} f(x_1,y)\! -\!{6 \over 5} g(x_1,y)\right) \sum_{i=0}^1 \dot{g}_{ii} dx^i dx^i
\!+  \!
\left(1\!+ \!{1 \over 5} f(x_1,y) \!-\!{6 \over 5} g(x_1,y) \right) \sum_{i=2}^4 \dot{g}_{ii} dx^i dx^i \nonumber \\
&& + \left(1+ {3 \over 5} f(x_1,y) -{6 \over 5} g(x_1,y) \right) \dot{g}_{yy} dy^2
+  \left(1+ {3 \over 5} f(x_1,y) +{3 \over 5} g(x_1,y) \right)  \dot{g}_{\Omega \Omega } d\Omega^2 \nonumber \\
e^{\phi} & = &  \left(1+ {9 \over 10} f(x_1,y) -{3 \over 5} g(x_1,y) \right)  e^{\dot \phi} ,\nonumber\\
C_0 & = & h(x_1,y).
\end{eqnarray}
By directly substituting these fluctuations into the supergravity
equations of motion,
\beq
R_{\mu \nu} - {1 \over 12}  H_{\mu \sigma \rho \lambda} {H_\nu}^{\sigma \rho \lambda} + {1 \over 144} H^2 g_{\mu \nu} = 0,
\eeq
we see that they satisfy the equations ($r = y/y_0$)
\begin{eqnarray}
(r^4-r^{-2}) f''(r) + (7 r^3 - r^{-3}) f'(r) + {9 M^2 \over 4 \pi^2
T^2} f(r) = 0, \nonumber\\
(r^4-r^{-2}) g''(r) + (7 r^3 - r^{-3}) g'(r) + \left({9 M^2 \over 4
\pi^2 T^2} g(r) - 72 r^2\right) = 0, \nonumber\\
(r^4-r^{-2}) h''(r) + (7 r^3 - 7 r^{-3})h'(r) + {9 M^2 \over 4 \pi
T^2} h(r) = 0.
\label{eq}
\end{eqnarray}

\subsection{Field/Operator Correspondence}
The string/gauge theory correspondence asserts that the string modes
$\varphi$ couple at the boundary to gauge invariant operators $\O$ of
the gauge theory.  The generating functional for the correlation
function of $\O$ is the string partition function evaluated with
specified boundary values $\varphi_0$ of the string fields.  When the
supergravity description is applicable we have
\beq
\langle e^{-\int  d^4x \varphi_0(x) \O(x)} \rangle = e^{-I_{SG}}(\varphi_0)
\label{cor}
\comma
\eeq
where $I_{SG}$ is the supergravity action.

The supergravity fluctuations $f,g,h$ couple to operators on the
D-brane world volume in flat background. For small 't Hooft parameter,
this coupling can be inferred from the form of the Born-Infeld
action. These same operators, in the strong coupling limit, are
described as ``sources at infinity'' in the dual AdS description.

The Born-Infeld Action describing the D4 brane worldvolume theory is
given by :
\beq
I_{BI} = \tr \T_4 \int d^5x\, \left( e^{-\phi} \sqrt{\det(G_{\mu \nu} +
{\cal F}_{\mu \nu} + G^{ij}\partial_\mu X_i \partial_\nu X_j)} + {1 \over 8}
\epsilon^{\mu \nu \sigma \rho \kappa} C_\mu {\cal F}_{\nu \sigma}
{\cal F}_{\rho \kappa}\right)
\comma
\eeq
where $i = 5 \ldots 9$ and  $\T_4$ is the D4 brane tension.

Dimensionally reducing along the $t$ direction on a circle with period
$1/T$ leads to
\beq I_{BI} = \tr {\T_4 \over T} \int d^4x\, \left(e^{-\phi}
\sqrt{G_{00}} \sqrt{\det(G_{\mu \nu}  + {\cal F}_{\mu \nu} + G^{ij}
\partial_\mu X_i \partial_\nu X_j)} + {1 \over 8} \epsilon^{0 \mu \nu
\sigma \rho} C_0 {\cal F}_{\mu \nu} {\cal F}_{\sigma \rho}\right)
\comma
\label{BI}
\eeq
where now, $i = 0,\ 5 \ldots 9$.  In order to restore the canonical
dimensions of Yang-Mills fields, we rescale $F$ and $X$:
\beq {\cal F} = \T^{-1} F, \qquad X = \T^{-1} \Phi \comma \eeq
where $\T^{-1} = 2 \pi l_s^2$.

Expanding the action (\ref{BI}) to quadratic order gives
\beq I = \tr \int d^4 x\, \left( {\T_4 \over 4 \T^2 T } e^{-\phi}
\sqrt{G_{00}} \sqrt{\det G_{\mu \nu}} F^2 + {\T_4 \over 8 \T^2 T} \epsilon^{0 \mu \nu \lambda \sigma} C_0 F_{\mu \nu} F_{\lambda
\sigma} \right)
\label{I}
\comma \eeq 
where we dropped the scalars since they are expected to decouple from
the dynamics in the far infra-red. Using \cite{CT97}
\beq
\T_p = g^{-1} (2 \pi)^{(1-p)/2} \T^{(p+1)/2}
\comma
\eeq
we get 
\beq {\T_4 \over \T^2 T} = {1 \over (2 \pi)^2 g l_s T} = {N \over 2
\pi \lambda} \comma \eeq
where in the last equality, we set $2 \pi g_s l_s = (\lambda/N)
T^{-1}$ which follows from M theory/Type IIA duality.  Before
substituting these fluctuations into (\ref{I}), note that the metric
fluctuation due to field $f(x_1,y)$ is polarized along $(3 (dx^1)^2 +
\sum_{i=2}^4 (dx^i)^2)$ which is not isotropic in the 1234 plane. One
could however think of this as contribution from two pieces $(
\sum_{i=1}^4 (dx^i)^2)$ and $(dx^1)^2$.  The second term must then
couple to $T_{11}(p_1)$ component of the stress-energy tensor which
vanishes due to conservation law $p^\mu T_{\mu \nu}=0$. We will
therefore think of fluctuation $f(x_1,y)$ as giving rise to a
fluctuation of the metric of the form
$$ds^2 = \left(1+ {1 \over 5} f(x_1,y) -{6 \over 5} g(x_1,y) \right)
\sum_{i=1}^4 \dot{g}_{ii} dx^i dx^i $$ 
along the 1234 plane. Substituting these fluctuations into (\ref{I}),
we find that $f(x_1,y)$, $g(x_1,y)$, and $h(x_1,y)$ couple to the world volume
operators via
\beq I = \tr \int d^4 x\, \left( {1 \over 4}\left({N \over 2 \pi
\lambda} \right) \left(1 - {3 \over 5} f(x_1,y)\right)  F^2 + {1 \over 8} \left({N
\over 2 \pi \lambda} \right)h(x_1,y) F_{\mu \nu} F_{\lambda \sigma}
\epsilon^{\mu \nu \lambda \sigma}\right) \stop
\label{I2}
\eeq
Note that all dependence on $g(x_1,y)$ canceled out to this
order. Indeed, the AdS/SCFT correspondence in six dimensions imply in
view of (\ref{eq}) that $g(x_1,y)$ couples in six dimensions to an
operator of dimension 12 in the (0,2) theory.  This operator is $H^4$
where $H=dB$, which reduces to a dimension 8 operator in the four
dimensional theory in the low energy limit. Comparing (\ref{I2}) to
the Yang-Mills action
\beq I_{YM} = \tr \int d^4x\,\left( {1 \over 4 g_{YM}^2} F^2 + {\theta
\over 16 \pi^2} F \tilde{F} \right) \comma
\label{YM}
\eeq
we get the relations
\beq g_{YM}^2 = {2 \pi \lambda \over N}, \qquad {\theta \over 16
\pi^2} = {1 \over 4} \left({N \over 2 \pi \lambda} \right) h(x_1,y) \stop
\eeq
The supergravity fields $f(x_1,y)$ and $h(x_1,y)$ couple to the operators
\beq \O_4 = {1 \over 4}\left({N \over 2 \pi \lambda}\right) {3 \over
5} \tr F^2, \qquad \tilde{\O}_4 = {1 \over 4} \left({N \over 2 \pi
\lambda}\right) \tr F \tilde{F}
\label{op}
\stop
\eeq

\section{Mass Spectrum}

The glueball masses in QCD can be obtained by computing correlation
functions of gauge invariant local operators or the Wilson loops, and
looking for particle poles. Following \cite{GKP98,WittenAdS}
correlation functions of local operators are related at large $N$ and
large $g_{YM}^2N$ to tree level amplitudes of supergravity.  We will
consider the two point functions of the operators (\ref{op}).  The
correlation function of $\O$ takes the form
\beq \langle \O(x) \O(y) \rangle = \sum_i c_i e^{-M_i|x-y|} \comma
\label{corr}
\eeq 
where $\O$ refers to (\ref{op}), and $M_i$ are the corresponding
glueball masses.  Therefore the spectrum of scalar
glueballs\footnote{In the following we will use the notation $J^{PC}$
for the glueballs, where $J$ is the glueball spin, and $P$, $C$ refer
to the parity and charge conjugation quantum numbers respectively.}
$0^{++}$ and $0^{-+}$ can be obtained by finding the normalizable  solutions to
(\ref{eq}) for $f(x_1,y)$ and $h(x_1,y)$ respectively.
The results are summarized in table
\ref{tab1}, where the masses are given in units of $T$.  We also
included the spectrum associated with the field $g(x_1,y)$.  

\begin{table}
\centerline{\begin{tabular}{|c|c|c|c|c|c|c|c|}  \hline
&$J^{PC}$&1&2&3&4&5&6 \\ \hline
$f(x_1,y)$ & $0^{++}$ & 9.85 & 15.6 & 21.2 & 26.7 & 32.2 & 37.7 \\ \hline
$g(x_1,y)$ & $0^{++}$ & 22.5 & 28.8 & 34.9 & 40.7 & 46.5 & 52.1 \\ \hline
$h(x_1,y)$ & $0^{-+}$ & 11.8 & 17.8 & 23.5 & 34.6 & 34.6 & 40.1 \\ \hline
\end{tabular}}
\caption{Mass spectrum in units of $T$ for $0^{++}$ and $0^{-+}$
glueballs corresponding to normalizable modes of fluctuations $f(x_1,y)$,
$g(x_1,y)$, and $h(x_1,y)$.\label{tab1}}
\end{table}

We can compare the ratio of masses of the lowest glueball states
$0^{-+}$ and $0^{++}$ with the lattice results \cite{Teper97}
\beqar
\left(\frac{M_{0^{-+}}}{M_{0^{++}}}\right)_{{\rm supergravity}}&= &1.20 \non\\
\left(\frac{M_{0^{-+}}}{M_{0^{++}}}\right)_{{\rm lattice~~~~~~}}& =&
1.36 \pm 0.32
\comma
\eeqar
and the results are in good agreement.

The spectrum of masses related to $g(x_1,y)$ is  higher than the spectrum
of masses related to $f(x_1,y)$ since it corresponds to a higher
dimensional operator.  Unlike QCD where different operators with the
same quantum numbers give the same mass spectrum, in the supergravity
limit the corresponding mass spectrum is different.  Since a $\tr F^4$
operator couples on the worldvolume theory to both $f$ and $g$ modes
we expect that in the QCD limit the spectrum of masses associated with
$g$ will decouple in order for $\tr F^2$ and $\tr F^4$ to give the
same spectrum.

Let us make some comments on the results.  In comparison to
\cite{COOT98} the masses of $0^{-+}$ are lower.  This is due to the
fact that in \cite{COOT98} the computation of the spectrum of $0^{-+}$
was done using the RR 3-form which couples to an operator with
$0^{-+}$ quantum numbers but of higher dimension than $\tr F
\tilde{F}$.  An example of that is the $0^{++}$ spectrum derived above
using the $f$ and $g$ modes.  Also, the masses of $0^{++}$ are not in
agreement with \cite{COOT98}.  This is because the supergravity
equation used in \cite{COOT98} to compute the $0^{++}$ mass spectrum
does not precisely correspond to $\tr F^2$ which is the lowest
dimension glueball operator with $0^{++}$ quantum numbers.

Finally, in order to compute the coefficients $c_i$ in (\ref{corr})
one has to evaluate the action for the supergravity fields requiring
that they approach a boundary configuration at infinity. This is a
procedure that can be done numerically, but will not be reported here.

\section{Topological Susceptibility}
The topological susceptibility $\chi_t$ is defined
by 
\beq
\chi_t= \frac{1}{(16\pi^2)^2} 
\int d^4 x \langle \tr F \tilde F(x)  \tr F \tilde F(0) \rangle
\stop
\label{topsus}
\eeq
The topological susceptibility measures the fluctuations of the
topological charge of the vacuum.  At large $N$ the Witten-Veneziano
formula \cite{Witteneta, Veneziano} relates the mass $m_{\eta'}$ to
the topological susceptibility of Yang-Mills theory without quarks
\beq m_{\eta'}^2 = \frac{4N_f}{f_{\pi}^2} \chi_t \stop
\label{mass}
\eeq

In this section we will compute the topological susceptibility in the
supergravity description.  The discussion is closely related to the
study of the $\theta$ dependence of the vacuum energy in
\cite{Wittentheta}.  Working in Type IIA supergravity, the action of
RR 1-form $C_{\mu}$ is given by
\beq
I = {1 \over 2 \kappa_{10}^2} \int d^{10} x \sqrt{g} {1 \over 4}
(\partial_{\mu} C_{\nu} - \partial_{\nu} C_{\mu}) (\partial_{\mu'}
C_{\nu'} - \partial_{\nu'} C_{\mu'}) g^{\mu \mu'} g^{\nu \nu'}
\stop
\eeq
For  the fluctuation $C_t = h(x_1,y)$, the action simplifies to 
\beq I = {1 \over 2 \kappa_{10}^2} \int d^{10} x \sqrt{g} {1 \over 2}
(\partial_{\mu} h(x_1,y)) (\partial_{\mu'} h(x_1,y)) g^{\mu \mu'} g^{tt} \stop
\eeq
The solution to the corresponding equation of motion (\ref{eq}) which
has dependence only in the $y$ coordinate is
\beq
h(y) = h^\infty (1 - {y_0^6 \over y^6})
\stop
\eeq
Plugging this into the action gives
\beq
I = {1 \over 2 \kappa_{10}^2} \int d^{10}x \sqrt{g} {1 \over 2}
g^{yy}g^{tt}(\partial_y h(y))^2 = 
 {h^\infty}^2 {2 \pi^2 y_0^6 \over \kappa_{10}^2 L^3 T} dx^4
\stop
\eeq
Using (\ref{conv}), (\ref{conv1}), dropping the volume factor and
differentiating twice with respect to $h^\infty$ we get
\beq \int d x^4 \langle \tilde{\O}_4(x) \tilde{\O}_4(0) \rangle = {128
\lambda N^2 \pi^3 T^4 \over 729} \stop \eeq
Recalling (\ref{op}) and (\ref{topsus}) we get the topological
susceptibility
\beq \chi_t = \int d x^4\left({\lambda \over 2 \pi N}\right)^2 \langle
\tilde{\O}_4(x) \tilde{\O}_4(0) \rangle = {32 \lambda^3 \pi T^4 \over
729} \stop \eeq

\section{Gluon condensate}
The gluon condensate $\langle {1 \over 4 g_{YM}^2} F^2 \rangle$ plays
an important role in the applications of Wilson OPE to the study of
strong coupling phenomena in QCD via the condensate expansion.  In
this section we will compute the gluon condensate in the supergravity
description.

According to the supergravity/gauge theory correspondence (\ref{cor}),
the one point function of an operator corresponds to the first
variation of the supergravity action. This quantity usually vanishes
identically by the equation of motion. In asymptotically anti
de Sitter spaces, however, there is a subtlety. The first variation is
only required to vanish up to a total derivative term. Since
asymptotically anti de Sitter space has a time-like boundary at
infinity, one must carefully examine the possible boundary
contribution. Indeed, the one point function of $F^2$ operator,
corresponding to first variation of the action with respect to the
field $f(x_1,y)$, will turn out to give a non-vanishing contribution.

This result can be anticipated without any detailed computation. The
non-extremal M5-branes has non-vanishing action density 
$I$ (which is
related to free energy $F$ by $I = T^{-1} F$)
\beq
I =  2^6 3^{-7} \pi^3 N^3 V_5 T^5  ={64 \over 2187} \pi^3 N^3 V_5 T^5
\stop
\eeq
When the M5-brane is wrapped on a circle of period
$(\lambda/N)T^{-1}$, we expect the corresponding free energy to be
given simply by making the substitution $V_5 \rightarrow V_4 T^{-1}
\lambda/ N$ which leads to an expression for the free energy of
non-extremal D4-branes:
\beq I = 2^6 3^{-7} \pi^3 \lambda N^2 V_4 T^4 ={64 \over 2187} \pi^3
\lambda N^2V_4 T^4 \stop \eeq

The gluon condensate operator can be gotten by acting on the
Born-Infeld action with $ \lambda {\partial \over \partial
\lambda}$. This suggests that one can act on $I$ with the same
operator to infer the value of $\langle {1 \over 4 g_{YM}^2} F^2
\rangle$ in the dual formulation:
\beq
\langle {1 \over 4 g_{YM}^2} F^2 \rangle = \lambda {\partial \over
\partial \lambda} I = {64 \over 2187} \pi^3 \lambda N^2V_4 T^4
\stop
\eeq

\section{Comparison}

We summarize the results obtained using the supergravity
description:\footnote{For completeness we quote also the result for
the QCD tension \cite{BISY98}. This result is corrected by a factor of $\pi^3$
compared to a previous version of this paper
due to a numerical error in a previous version of \cite{BISY98}.}
\begin{eqnarray}
M_{0^{++}} & = &  9.86 T\non\\
M_{0^{-+}} & = &  11.8 T\non\\
{\rm QCD\ string\ tension}& = & {16 \pi^2\over 27} \lambda T^2\non\\
\chi_t & = & {32  \pi \over 729}  \lambda^3 T^4 \non\\
\langle {1 \over 4 g_{YM}^2}\tr F^2(0) \rangle & = & {64 \pi^3 \over
2187} N^2 \lambda T^4
\stop
\label{SG}
\end{eqnarray}
These results are the leading in the strong coupling expansion of the
gauge theory.

The lattice QCD results are \cite{Teper97}\footnote{The lattice
results have systematic and statistical errors that we do not quote.}

\begin{eqnarray}
M_{0^{++}} & \simeq &  1610 \MeV \non\\
M_{0^{-+}} & \simeq &  2190 \MeV \non\\
{\rm QCD\ string\ tension}& \simeq & (440 \MeV)^2 \non\\
\chi_t & \simeq & (180 \MeV)^4 \non\\
\langle {1 \over 4 g_{YM}^2}\tr F^2(0) \rangle & \simeq &  {N^2 \over  \lambda} (400 \MeV)^4 
\stop
\end{eqnarray}
The entry for the gluon condensate is derived by scaling the $N=3$
result of \cite{gi} by a factor of $N/3$.
\beq
{N \over 3} 0.1 {\rm GeV}^4 =  \langle {\alpha_s \over \pi} G^2  \rangle
= \langle {1 \over 2 \pi^2}  F^2 \rangle 
\stop
\eeq
Note that plugging the phenomenological values $N_f=3, N=3, m_{\eta'}
\sim 1 \GeV,f_{\pi} \sim 0.1 \GeV$ in (\ref{mass}) leads to a
prediction $\chi_t \sim (180 \MeV)^4$, in agreement with the lattice
simulation.

The supergravity results depend on two parameters, $\lambda$ and
$T$. This should be thought of as the leading $1/\lambda$ asymptotic
behavior of the full string theory expression $(F(\lambda) T)^d$ where
$d$ is the dimension of the observable. Ideally, the comparison with
the lattice QCD result should be made in the limit where $\lambda$ is
taken to be small and $T$ is taken to be large. In this range of
parameters, one can trust perturbative QCD, which tells us that
$F(\lambda)$ goes as
\beq
F(\lambda) = C e^{-{12 \pi \over 11 \lambda}} \stop
\eeq
where $C$ is some numerical coefficient.  The genuine QCD data is
encoded in the ratio of this numerical coefficient $C$ between
different observables. Unfortunately, this is precisely the
information that is not accessible via the supergravity
approximation. In the absence of fully string theoretic formulation of
this system that allows one to probe the small $\lambda$ region, one
must contend oneself with qualitative comparisons.  To facilitate such
a comparison at a qualitative level, it is useful to go to the
so-called ``correspondence point'' in the $\lambda$-$T$ parameter
space. This is the point where dual descriptions of the theory crosses
over from one to the other.  As such, one can not trust neither
description at a quantitative level, but one expects the results to
match up to numerical factors of order one.  One can estimate the
cross-over point based on natural scale of our problem: $\lambda
\approx {12 \pi \over 11}$ and $T \approx \Lambda_{{\rm QCD}} \approx
200 \MeV$. With this choice of $\lambda$ and $T$, the ratio of
observables in units of energy comes out as follows:
\begin{eqnarray*}
{M_{0^{++}}^{{\rm SUGRA}}  \over M_{0^{++}}^{{\rm Lattice}}} & = & 1.2 \\
{M_{0^{-+}}^{{\rm SUGRA}}  \over M_{0^{-+}}^{{\rm Lattice}}} & = & 1.1 \\
\left({{({\rm QCD\ string\ tension})}^{{\rm SUGRA}}  \over {({\rm QCD\ string\ tension})}^{{\rm Lattice}}}\right)^{1/2} & = & 2.0\\
\left({\chi_t^{{\rm SUGRA}}  \over \chi_t^{{\rm Lattice}}}\right)^{1/4} & = & 1.7 \\
\left({{({\rm Gluon\ condensate})}^{{\rm SUGRA}}  \over {({\rm Gluon\ condensate})}^{{\rm Lattice}}}\right)^{1/4} & = & 0.9
\end{eqnarray*}
It is encouraging to find that these ratio are indeed of order unity 
\footnote{The numerical value for the ratio of the QCD string tensions is modified
compared to a previous version of this paper
in view of the change in the numerical value
of the QCD string tension in (\ref{SG}).}.

It is curious to note for comparison the known results in strong
coupling lattice QCD \cite{Creutz}:
\begin{eqnarray}
M_{0^{++}} & = &  -4 \log(g_{YM}^2 N) a^{-1}\non\\
{\rm QCD\ string\ tension}& = & \log(g_{YM}^2 N) a^{-2} 
\stop
\label{lattice}
\end{eqnarray}
They reflect the fact that the $\alpha'$ expansion is not the same as
the lattice strong coupling expansion. In particular, while in the
supergravity description the ratio of $M_{0^{++}}$ to the QCD string
tension goes to zero, it diverges in the strong coupling lattice QCD.

\section{Summary and Discussion}

In this paper we computed various physical quantities which are of
importance in QCD dynamics.  The computations were performed in the
framework of supergravity.  Since the supergravity description is
believed to provide the leading contribution in the strong coupling
expansion of the gauge theory, our results should be viewed as a first
step in understanding the QCD dynamics via the string/gauge theory
correspondence.  Clearly, a complete solution requires a detailed
knowledge of the string dynamics with R-R background.

The dual supergravity description corresponds to taking the bare
coupling to infinity. This is the opposite limit to continuum QCD
where the bare coupling is taken to zero.  Therefore it is not
possible, for instance, to exhibit asymptotic freedom via the
supergravity description.  In analogy with strong coupling lattice
QCD, it is therefore far from clear what properties of continuum QCD
should be captured by supergravity.  The supergravity description
includes also Kaluza-Klein modes that do not have corresponding
degrees of freedom in QCD.  They reflect the physics of the higher
dimensions and should decouple in the small bare coupling limit.
Since the background is singular in this limit, their decoupling is a
subtle issue.

We can repeat the analysis for QCD$_3$.  Here the procedure is to
compactify on a circle the D3 branes of type IIB string theory and
introduce anti periodic boundary conditions for the fermions
\cite{WittenAdsThermal}.  We find\footnote{For
completeness we quote also the results for the $0^{++}$ mass
\cite{COOT98} and the QCD tension \cite{BISY98}. The spectrum in
\cite{COOT98} was quoted in units of $b$, which in AdS$_{n+1}$
Schwarzschild space, is related to $T$ by $b = 4 \pi T/n$.}
\begin{eqnarray}
M_{0^{++}} & = &  10.7 T \non\\
{\rm QCD\ string\ tension}& = & \pi\sqrt{{1 \over 2}g_{YM3}^2 N T^3} \non \\
\langle {1 \over 4 g_{YM3}^2}\tr F^2(0) \rangle & = & {\pi^2 \over 8} N^2 T^3
\stop
\end{eqnarray}
The gluon condensate in QCD$_3$  has not been computed on the lattice yet.

\section*{Acknowledgement}

We are grateful to 
O.~Aharony, 
S.~Carroll,
C.~Csaki,
D.~Gross,
G.~Horowitz, 
H.~Ooguri,
A.~Peet,
J.~Polchinski,
M.~Srednicki,
M.~Teper,
J.~Terning,
A.~Tseytlin, 
and
E.~Witten
for useful discussions. The work of AH was supported in part by the
National Science Foundation under Grant No. PHY94-07194.

\begingroup\raggedright\endgroup

\end{document}